\documentclass[10pt, conference, letterpaper ]{IEEEconf}

\usepackage{lineno}
\usepackage{amssymb}
\usepackage{adjustbox}
\usepackage{epsfig}
\usepackage{graphicx}
\usepackage{capt-of}
\usepackage{graphics}
\usepackage{float}
\usepackage{bbm}
\usepackage{subfiles}
\usepackage{physics}
\usepackage{multirow}
\usepackage{enumerate}
\usepackage{booktabs} 
\usepackage{tikz}
\usepackage[theorems,skins]{tcolorbox}

\usepackage{url}
\usepackage{etoolbox}
\usepackage{bbm}
\usepackage{amsmath}
\usepackage{caption}
\usepackage{empheq}

\newcommand{\DetailK}[2]{#1}
\newcommand{\psiL}{\psi_a^L}
\newcommand{\baL}{\beta_a^L}
\newcommand{\taL}{\theta_a^L}

\newcommand{\tL}{\theta^L}
\newcommand{\bL}{\beta^L}

\newcommand{\nto}{\nrightarrow}
 
\newcommand{\n}{\eta}
\newcommand{\N}{\mathcal{N}}

\newcommand{\I}{I_{\theta/\psi}}

\newcommand{\eop}{\hfill{$\blacksquare$}}
 
\newtheorem{theorem}{Theorem}

\newtheorem{lemma}{Lemma}

\newtheorem{remark}{Remarks}
\newcommand{\Detail}[1]{}  

\newcommand{\Remove}[1]{}  
\newcommand{\tobeputagain}[1]{}

\usepackage{bm}
\usepackage{framed}
\usepackage{ifthen}
\newcommand{\up}{\uparrow}

\newcommand{\beq}{\begin{eqnarray*}}
\newcommand{\eeq}{\end{eqnarray*}}

\usepackage{csquotes}

\bibliography{references}
\begin{document}
 \title{Branching Process with Attack: Viral Competing Markets} 
 \author{Khushboo Agarwal, Veeraruna Kavitha \\
IEOR, Indian Institute of Technology Bombay, India}
\maketitle

\begin{abstract}
The marked increase in advertisements over online social networks (OSNs) necessitates the study of content propagation. We analyse the viral markets with content providers competing for the propagation of similar posts over OSNs. Towards this, we required a new variant of the branching process (BP), which we named as ``Branching process with attack"; the entities upon wake up attempt to attack and acquire the opposite population; furthermore, each entity produces its offsprings as is usually considered in BPs.

In addition to providing expressions for the growth rates of individual posts, dichotomy etc., we explore the co-existence of posts; can the competing content spread and explode (number of unread/live copies of both posts grow significantly with time) simultaneously over the network? We prove that either one or both populations/posts get extinct or the populations settle to a unique co-existence equilibrium and derive the corresponding asymptotic ratios of the two populations/posts. Our analysis applies to large population networks focusing on mass behaviour, rather than micro details. Our study provides insights into two crucial design aspects, the number of seed users and the quality of the post. 

\end{abstract}
 

\section{Introduction}

Social media is universally prevalent today; the shared content is shared (again), liked or dis-liked by the users and thus may get viral over the network. This gives the content providers (CPs) an opportunity to share the product information with customers and  cheaply amplify their growth. This strategy is well known as ``Viral Marketing'' (e.g., \cite{dhounchak2017viral, iribarren2011branching, van2010viral}).  Variety of content propagates simultaneously through such platforms, and enjoy the resultant benefits; however, they face strict competition due to competing contents. 

 In \cite{dhounchak2017viral}, a timeline structure holding the content is considered to study the competing content dissemination over online social networks (OSNs). Here the competition  was due to the placement of competing content at various levels of the timeline, but the paper did not consider the aspect that a user would choose one among the competing contents. 
When a user, already shared with one content (say content A), is also shared with another content (say content B) which is in direct competition with the former,  it may choose the latter post. Thus one can say that `content A is attacked and acquired by content B'. 
We study such systems and refer them as \underline{\textit{viral competing markets}}. Each CP has to decide on two factors: (i)  the initial set of users (called seed users) for whom it may have to pay to share its content; (ii) the quality of its post. Some of the key questions that we explore are: 
\begin{enumerate}
    \item At what growth rate does each content propagate?
    \item With what probability does a given content get extinct?
    \item Can the posts of competitors  co-exist over the network? 
    \item How should a CP trade-off between investing on  seed users and designing an attractive post?
\end{enumerate}
Some of these questions were investigated previously (e.g., \cite{beutel2012interacting, deijfen2016winner, prakash2012winner, zhou2019cost}), but the presence of competition poses new challenges and provides a new perspective towards viral markets.

\textit{Our approach and contributions:} There are several approaches for studying content propagation over the network. A set of literature focuses on micro details, like degree and particularity (basically, connections) of the users in the network for their designs   and uses tools like random graphs (e.g., \cite{deijfen2016winner, zhou2019cost}) or epidemiology based models (e.g. \cite{beutel2012interacting, prakash2012winner}) for the study. However, these models can not capture some important aspects related to content propagation like virality, i.e., the explosion of (the number of shares of) content over the network. \textit{We consider large   population networks, where  all users are typically the same (in stochastic sense), and focus on macro-details related to content propagation.} Branching processes (BPs) are a great choice for such a study (e.g., \cite{dhounchak2017viral, iribarren2011branching, van2010viral}). It provides analysis for growth patterns, extinction probability and other measures,  and captures phenomenon which are specific to content propagation and virality. Some existing research considers competing contents over the network using techniques other than BPs \cite{beutel2012interacting, deijfen2016winner, prakash2012winner}; however, those models are inadequate for studying \textit{co-virality of the posts}, i.e., concurrent explosion of the posts. We aim to understand these phenomenon in this paper.

There is a vast literature that studies different variants of BPs (e.g., \cite{athreya2006branching,coffey1991galton, klebaner1984population}, to name a few). The prey-predator type branching process (\cite{coffey1991galton}) is the only BP that can capture some aspects of competition. Here, one  population  (predator) attacks, while the other population (prey) progresses like a standard BP. They do not consider ``double-sided attack'' and ``acquisition''. Further they consider discrete-time framework, while the viral competing markets evolve  continuously over time. Thus, the existing branching models  are insufficient to mimic important aspects of viral competing markets.

We introduce a new variant of Multi-type branching process, named as \underline{\textit{Branching process with attack} (BPA)}.   In standard BP, upon wake up, each entity produces a random number of its own offsprings before dying; in our model, it also attacks and captures the opposite population. We model the unread copies of each post as one population.  Our models (and other variants like \cite{coffey1991galton, klebaner1984population}), do not satisfy the well-known self-similarity property of BPs, i.e., one can not extract a sub-tree from any node which replicates the behaviour of the original tree, this significantly complicates the study.


We contribute towards both branching process and viral marketing literature. In addition to studying growth rates, dichotomy etc., we 
provide a very distinct result in contrast to ``winner takes it all" (studied in \cite{prakash2012winner, deijfen2016winner} and others). We prove using \textit{stochastic approximation techniques} that either one or both posts get extinct, or the limit proportions converge to a unique fixed ratio which is a function of mean offsprings and   attack capacity. The third possibility is a rare phenomenon and occurs when mean number of offsprings and/or initial population is high (as in large OSNs). Our results suggest that a CP should invest more on the number of seed users than on improving the quality of its post. 


Our BPs can also be used for various other applications, e.g., election campaigns, propagation of content related to awareness programs, and biological phenomenon.


\section{Viral Competing Markets}\label{sec_problem_desc}
We consider content propagation over online social networks by two  content providers (CPs)   competing for  similar kind of product/service. The main goal of each CP is to reach out to more audience through viral marketing; a certain fraction of the users receiving the post may provide business to the CP. Each CP initially posts its content to few selected users   and these are referred to as seed users.  
Whenever a receiver visits the application (referred to as OSN, Online social network) over which the post has been shared,  it views the post. We refer  the time instance at which  the user views the post as its ``wake-up'' time. After viewing, the user shares it with some or all of its contacts/friends. The number of shares depend upon the extent to which the user liked the post. We assume that the number of friends ($ F$) of a typical user is random which is independently and identically distributed (IID) across the users.  Our main aim is to analyze the propagation of such competing content over OSNs.  

\textbf{\textit{Branching process:}}
As is usually done in viral marketing literature (e.g., \cite{dhounchak2017viral,  iribarren2011branching, van2010viral}), we are modeling this propagation using an appropriate branching process (BP). We detail the dynamics of the problem, along side, describing the BP based modeling details. Towards the end, we would observe that the branching process modeling the propagation of competing content is very different from the models considered in literature.

$\bullet$
We would refer the two CPs respectively  by $x$ and $y$ type CPs.
If the user receives only one kind of post (say of $x$-CP), at the time it views the post/it wakes-up, we refer it as $x$-type post. In this case, the user would obviously  share (if at all) only $x$-type post to some of its friends. Once viewed, the user would not read the post again, we then say that the number of unread $x$-type users/posts have reduced by one, and are increased by the number of shares; this is exactly like a death in an appropriate BP, after producing random number of offsprings (shares here). 

 
$\bullet$
If the user receives both the posts at its wake-up, (due to competing nature of the two posts) it may chose one among the two to  share; we refer such a post/user as $x$ or $y$-type respectively depending upon its choice. It is also possible that it shares to none.  Initially the user must have been shared (say) $x$-post, we then refer it as $x$-type. But when another friend forwards the $y$-type post before the user wakes-up, then the user gets converted to $y$-type if it prefers $y$-post over the $x$-post at wake-up. {\it It is this conversion, which we refer  as attack (by $y$-type),} that makes the underlying branching process very different from the ones studied in the literature.  As before the corresponding unread posts are reduced by one, once the user views (and shares) them. Observe  that a user not sharing the post to any of its friends, is equivalent to zero offsprings. 

$\bullet$ If the user is shared with multiple copies of the same post, then the user would  consider only the latest share. This effect is considered by appropriate modelling of offspring distribution (details in section \ref{sec_viral_markets}).

The posts propagate from one user to another in the above described manner. As mentioned before, there is a possibility of the rival post reaching the user before it views the first received post. If the user prefers to share the rival post, then one can say that ``the rival post attacked and acquired   the first post''. These two features of ``attack" and ``acquisition" are the novel features of what we call as {\it branching process with attack (BPA).} We would provide a precise description of the process, analyse it in the next two sections before returning to viral marketing application in section \ref{sec_viral_markets}.

\section{Branching process with Attack (BPA)}\label{sec_BPA}
Consider a branching process (BP) with two types of population, $x \mbox{ and } y$. We consider Markovian BPs;   any particle wakes-up after an exponentially distributed time with parameter $\lambda$, and, produces random number of offsprings before dying. Our  branching processes with attack (BPA) also includes the attack by rival types.

Let $X(t)$, $Y(t)$ represent the size of the population of respective types at time $t$. If $x$-type individual wakes-up at time $t$, it produces  $\xi_x$ number of offsprings of its own type and attacks $\zeta_{xy}$ number of individuals of $y$-type; one individual can attack at maximum $\xi_{xy}$ number and the $y$-population opposes the attack with probability $p_{xy}$, in all, we assume that ($Bin$ is a Binomial random variable) 
\begin{equation}\label{eq_zeta}
    \zeta_{xy}(Y(t)) = Bin(\min\{\xi_{xy}, Y(t)\}, p_{xy}).
\end{equation}
The $y$-population  also evolves in exactly the similar manner. 
The offsprings and maximum possible attacks, $\{\xi_x\}$  and $\{\xi_{xy} \}$, are IID across $x$-type population, similar is the case with $\{\xi_y\}$ and $\{\xi_{yx}\}$. Our assumptions are:

\begin{enumerate}[{\bf A}.1]
    \item We assume \textit{super critical} conditions and also finite first  moments, i.e., 
for each $i, j \in \{x, y\}, i \neq j$:
$$1 < m_i := E(\xi_{i}) < \infty, \ m_{ij} :=  E(\xi_{ij})    < \infty.$$
    \item We assume finite second  moments, i.e., 
for each $i, j \in \{x, y\}, i \neq j$:
$E(\xi_{i}^2) < \infty, \ E(\xi_{ij}^2)    < \infty.$ 
\end{enumerate}
For some of our results, we do not require  \textbf{A}.2; we will mention   explicitly,  when required. 
The expected number of (successful) attacks depend upon the size:
\begin{eqnarray}
\label{Eqn_mxy}
    m_{xy} (y) &:= & E(\zeta_{xy}(y)) = E[Bin ( \min\{\xi_{xy}, y\}, p_{xy} ) ] ,\\
\label{Eqn_mxy_star}
    m_{xy}^* &:=& \lim_{y \to \infty} m_{xy}(y) .
\end{eqnarray}
Define $m_{yx} (x), m_{yx}^*$ in a similar way.
%
We also assume that $0 < P(\xi_{i} = 0) < 1$, for each $i \in \{x, y\}$, thus ensuring  a possibility of zero as well as higher number of  offsprings. Let us denote ($\mbox{for }  i, j \in \{x, y\}, \ \ i \neq j$): $\alpha_i := \lambda(m_i - 1) $.

Let $x_0, y_0$ be the (deterministic) initial population sizes of $x, y$ populations respectively. 

%
 
\vspace{-1mm}
\subsection{\textbf{Dynamics of BPA:}}
We analyse BPA by studying the respective population sizes at transition epochs, i.e., at time instances at which an individual wakes-up. We begin with some notations. Let us denote the event that $x$-type wakes up as $x\up$, and, similarly define $y\up$. Let $\tau_n$  represent the $n^{th}$ transition epoch\footnote{If the entire population gets extinct at $n^{th}$ epoch, we set $\tau_{k} :=\tau_n$ for all $k \geq n$. }.
Let $X_n := X(\tau_n^+)$ be the size of $x$-type population immediately after $\tau_n$. Similarly  define $Y_n$. Observe that the {\it time taken by the first individual to wake-up, ($\tau_{n+1}-\tau_n$), after the $n^{th}$ transition epoch is exponentially distributed with parameter  $\lambda(X_n + Y_n)$.}  One can summarize   BPA using the following description of the events at transition epochs. When an $x$-type individual wakes up ($x \up$), it produces $\xi_x$ offsprings, attacks $\zeta_{xy}$ number of $y$-type and dies, i.e., (\mbox{see } \eqref{eq_zeta}),

\vspace{-0.2cm}
{\small\begin{equation}\label{eqn_evolve}
\begin{aligned}
X_{n+1} = X_n - 1 + \xi_{x} + \zeta_{xy}(Y_n) ,  \ \ \
Y_{n+1} = Y_n - \zeta_{xy} (Y_n).
\end{aligned}
\end{equation}} 
Exactly similar transitions occur  
  when a $y$-type individual wakes up ($y \up$). 
Observe that the (successfully) attacked individuals are acquired by the attacking population. 

We will  immediately present a result about co-existence (both population survive),  that is true with  only \textbf{A}.1.

\subsection{\textbf{Non-coexistence under high-separation}}\label{sec_non_co_existence} 
Consider the scenario with high separation between the initial population sizes. Then, we have  (proof in  Appendix): 
\begin{theorem}\label{thrm_high_sep}
Assume $m_{xy}(1) >0$ and {\bf A}.1. For any $\epsilon > 0$ and $y_0$,  there exists a $x_{\epsilon, y_0} < \infty $ such that
\begin{align*}
     P ( Y\rightarrow 0 |  X_0 =  x_0, Y_0 = y_0) >  1 - \epsilon  \mbox{ for all } x_0 \geq x_{\epsilon, y_0} .
\end{align*} 
The same result is true when $x$ and $y$ are interchanged.  \eop
\end{theorem}

The above Theorem implies that given $y_0$ ($x_0$), there exists large enough $x_0$ (respectively $y_0)$ such that the population with the smaller initial size gets extinct with high probability,  i.e., {\it co-existence is not possible.}  Note that the above result is true irrespective of $\xi_x \stackrel{d}{=} \xi_y, \xi_x \stackrel{d}{<} \xi_y \mbox{ or } \xi_x \stackrel{d}{>} \xi_y$ (equality and inequality in stochastic sense), i.e., \textit{no matter which type of population produces higher mean number of offsprings, the    population with smaller initial size will get extinct with high probability.} It only  requires that the higher population is attacking (i.e., e.g., $m_{xy} (1) >0$). 

\section{Analysis of BPA}\label{sec_analysis_BPA}
We begin our analysis with the symmetric case where $\xi_x \stackrel{d}{=} \xi_y$ i.e., the two offspring distribution   are stochastically equal. Hence, the mean number of offspring   are  equal, let  $m_x = m_y := m$  and $\alpha := \lambda(m-1)$.  One can have asymmetric attacks, i.e., the distributions of $\xi_{xy}$ and $\xi_{yx}$ could be different. We consider asymmetric case (i.e., $\xi_x \stackrel{d}{>} \xi_y$) towards the end of this section.
Let $\mathcal{F}_n$ be the sigma-algebra generated by $X_n,Y_n,\tau_n$ for all $n$. {\it All the (sub/super) martingales  are with respect to  (w.r.t.) this filtration.}

\subsection{\textbf{Dichotomy}}\label{dichotomy} Let the total, maximum and minimum populations be defined respectively by $S_n = X_n + Y_n$,  $U_n = \max \{X_n, Y_n\}$ and $V_n = \min\{X_n, Y_n\}$. Under symmetric conditions, one can easily analyse some of these processes using the theory of standard BPs.
From \eqref{eqn_evolve},  irrespective of the type waking up, we have (under symmetric conditions, $\xi_y \stackrel{d}{=} \xi_{n+1} \stackrel{d}{=} \xi_x$):
\vspace{-0.1cm}
\begin{eqnarray}
S_{n+1} = S_n + \xi_{n+1} - 1 .\label{Eqn_Sn}
\end{eqnarray} 
Thus, the total population is the embedded chain corresponding to a standard single type BP \cite[Chapter 3]{athreya2006branching}).  Further note that   
$S_n/2 \le U_n \le S_n$. Hence we immediately have the following result (Proof of (c) in Appendix): 

\begin{theorem}\label{growth_total_pop} Assume \textbf{A}.1 and $\xi_x \stackrel{d}{=} \xi_y$.  Then, \newline
(a) The process $\{S_n e^{-\alpha \tau_n }\}_n$ is a non-negative martingale w.r.t.  sigma algebras $\{\mathcal{F}_n\}_n$, and converges almost surely (a.s.)  to a non-degenerate and integrable limit $W_s$. 
Further, 
\begin{equation}
\small{P(W_s = 0) = P(S_n    \to 0) =  (q^{*})^{(x_0 + y_0)} =  P(U_n  \to 0) ,} \label{Eqn_survive_prob}
\end{equation}
where $q^{*}$ is the unique solution of the fixed point equation $f(s) = s$ in $[0, 1)$, where $f(\cdot)$ is the PGF of $\xi_x$ or $\xi_y$. \newline
(b) $U_n$ exhibits dichotomy: {\small $P(U_n \to 0 \cup U_n \to \infty) = 1$}, and;\\
(c) $V_n$ exhibits dichotomy: {\small$ P(V_n \to 0 \cup \ \lim \sup V_n = \infty) = 1.$} 

\eop
\end{theorem}
The significance of this theorem is well understood in BP literature and we would explain the same in our context:\\ \textbf{i)} in the sample paths with  $W_s > 0$, the total population {\it grows at (time) asymptotic  rate $\alpha$}; for large enough $N$ and an appropriate $\epsilon > 0$, we have $ |W_s - S_n e^{-\alpha \tau_n} | <    \epsilon$ and so:
\begin{align}
\label{Eqn_growth_rates}
   (W_s-\epsilon) e^{\alpha \tau_n}  <   S_n <  (W_s + \epsilon )  e^{\alpha \tau_n}  \mbox{ for all }  n \ge N.
\end{align}
Observe here that $\tau_n \to \infty$ a.s. on sample paths in which at least one of the populations survive ($\tau_n$ is lower bounded by maximum among $n$ exponential random variables); \\
\textbf{ii)} the asymptotic growth rate of    $U_n$ is also given by $\alpha$; \\ \textbf{iii)} the {\it probability of extinction} of both $S_n, U_n$, i.e., the {\it probability that either population becomes zero for some $n < \infty$} is given by \eqref{Eqn_survive_prob};
\\
\textbf{iv)} further from \eqref{Eqn_survive_prob}, $U_n, S_n$ {\it exhibit dichotomy:} either they   have  asymptotic (exponential) growth  (whenever they survive, then $W_s > 0$) or they get extinct (then $W_s = 0$); and, \\
\textbf{v)} $V_n$ also exhibits dichotomy, i.e., either dies or grows to infinity (at least along a subsequence and at a rate not more than (exponential) $\alpha$).  

Consider the standard BP where none of the population types attack each other (i.e., $\xi_{xy} \equiv 0 \equiv \xi_{yx}$); we call such a process as \textbf{branching process with no attack (BPNA)}. In BPNA, under symmetric conditions, the total population grows like a single-type BP. 
In fact, it is easy to verify that the process corresponding to the total population is  same as that with attack, i.e.,  the phenomenon of attack does not change  the behavior of total population.  

\subsection{\textbf{Co-existence and limit proportions}}\label{sec_co_exist}
By Theorem \ref{thrm_high_sep}, co-existence is not possible if there is a huge gap in the initial population sizes. Here, we analyse the co-existence of populations in a more generic setting, using \underline{\textit{stochastic approximation (SA) techniques}} (e.g. \cite{kushner2003stochastic}). We also discuss the limit proportions. Towards this, let $\S_n $ represent the  sample mean formed by IID sequence of offsprings $\{\xi_n\}_n$ generated plus initial populations  (see \eqref{Eqn_Sn}):
\begin{center}
$\S_n = \frac{1}{n} \left (\sum_{k=1}^n (\xi_k -1) + x_0 + y_0 \right ).$ 
\end{center}
Observe that the total population ($\nu_e$ extinction epoch),
\begin{center}
    $S_n = n\S_n 1_{\{n < \nu_e\}}$, $\nu_e := \inf \{n : S_n = 0\}.$
\end{center}
Further, also observe (same is the case for $Y_n$):
\begin{equation}\label{Eqn_XnYnSnetc}
     X_n \le S_n \le n|\S_n| \mbox{ for all } n.  
\end{equation}
By strong law of large numbers,
  $\S_n \to m-1$ a.s., while $S_n/n \to m-1$ only in survival sample paths. These observations form the  main basis for the proceeding analysis.
Define $\Theta_n = [\psi_n, \theta_n, t_n]$, the ordered triplet respectively representing $S_n/n$, $X_n/n$ and $\sum_{k =0}^ {n-1}  1/(k+1)$.
We will show in the following that their evolution can be captured by a 3-dimensional stochastic approximation based scheme. 

We begin with some notations, required for this subsection. Let $H_n$ be the indicator that an $x$-type wakes-up ($x \up$) at the $n^{th}$-transition epoch and $H_n^c := 1-H_n$. Further, let $I_n := 1_{\{\theta_n < \psi_n\}}$, $J_n  := 1_{\{\psi_n > 0, \theta_n > 0\}}$, $K_n = 1_{\{ \psi_n > 0\}}$ and $\epsilon_n := 1/(n+1)$.
Then the evolution of $\Theta_n$ is given by the following  (see~\eqref{eqn_evolve}): 
\vspace{-0.25cm}
{
\begin{equation}\label{eq_stoch_approx_scheme}
\begin{aligned}
\psi_{n+1} &=  \psi_n + \epsilon_n \big (\xi_{n+1} - 1 - \psi_n \big )K_n,
\\
\theta_{n+1} &= \theta_n + \epsilon_n \bigg ( H_n \left (\xi_{n+1}-1+\zeta_{xy, n}(Y_n)I_n \right ) \\
&\hspace{2cm}-H_n^c\zeta_{yx, n}(X_n)I_n - \theta_n  \bigg ) J_n,\\
t_{n+1} &=  t_n  + \epsilon_n ,  \ \  \  \ \Theta_0 = [x_0 + y_0, x_0, 0],
\end{aligned}
\end{equation}} 
\vspace{-0.1cm}
and note $X_n = \n (t_n) \theta_n, Y_n = (\psi_n-\theta_n) \n (t_n)$, with  
\begin{center}
$\n (t) := \max\left  \{ n: \sum_{k=0}^{n-1}  \frac{1}{k+1} \le t \right \}.$
\end{center}

We analyse the above using the results of  \cite{kushner2003stochastic} and hence the notation specific (only) to this subsection closely matches  with its notation.
Define $L_n := [L_n^\psi, L_n^\theta, L_n^t]^T$,  where, 

\vspace{-0.1cm}
{\small
\begin{equation}
    \begin{aligned}
    L_n^\psi &= K_n(\xi_{n + 1} - 1 - \psi_n), \ \  L_n^t = 1, \mbox{\normalsize and } \\
    L_n^\theta &= \bigg( H_n\left(\xi_{n+1}-1+\zeta_{xy, n}(Y_n)I_n\right)  \\
    &\hspace{1.1cm}- H_n^c\zeta_{yx, n}(X_n)I_n - \theta_n \bigg)J_n .
\end{aligned}
\end{equation}
}
Writing \eqref{eq_stoch_approx_scheme} compactly,  $\Theta_{n+1} = \Theta_n + \epsilon_n L_n.$
The conditional expectation of $L_n$ with respect to ${\cal F}_n$ is given by:  

\vspace{-4mm}
{\small
\begin{align}
 E[L_n|\mathcal{F}_n] &= \bar{g}(\Theta_n), \mbox{\normalsize where, }  \\
\bar{g}^\psi (\Theta) &:=        1_{\{\psi >0\}}(m-1-\psi), \  \  \bar{g}^t (\Theta)    : = 1,  \ \mbox{\normalsize  and, }  \nonumber \\
\bar{g}^\theta (\Theta)  &: = \bigg(\frac{\theta}{\psi} \left (m-1+ m_{xy}\left(\left(\psi-\theta\right)\eta(t)\right)1_{\{\theta < \psi\}} \right)\nonumber \\
    &\hspace{-7mm} - \left(1-\frac{\theta}{\psi}\right)m_{yx}\left(\theta \eta(t)\right)1_{\{\theta < \psi\}} - \theta \bigg ) 1_{\{\psi > 0, \theta > 0\}} .
\nonumber
\end{align}}
With these definitions, the ODE  (Ordinary differential equation)
that can  approximate \eqref{eq_stoch_approx_scheme} is given by (see \cite{kushner2003stochastic}):
\begin{equation}\label{ODE}
    \dot{\psi} = \bar{g}^\psi (\Theta),  \ 
    \dot{\theta} = \bar{g}^\theta (\Theta), \mbox{\normalsize and } 
    \dot{t} = \bar{g}^t (\Theta).
\end{equation}
We will prove that the ODE indeed approximates  \eqref{eq_stoch_approx_scheme} and derive further   results mainly using \cite[Theorem 2.2, pp. 131]{kushner2003stochastic}
Since  ${\bar g}(\cdot)$ is measurable, and the  ODE is non-autonomous, the results can not be  applied directly. We provide the required justifications/modifications, identify the attractors and the domain of attraction of  the ODE  and finally derive the following result (proof and details in Appendix) under the additional assumption: 




\begin{enumerate}[{\bf A}.3]
    \item For some finite $\bar{y}, \bar{x}, \kappa_{ij}$, for $i, j \in \{x, y\}, i \neq j$, 
$$\kappa_{ij} \min\{j, {\bar j}\} \le m_{ij} (j) \le {\bar j} \kappa_{ij} = m_{ij}^*,$$
i.e., $m_{xy}(\cdot), m_{yx}(\cdot)$ are bounded by piece-wise linear functions.
\end{enumerate}
    


\begin{theorem}\label{thrm_co_exist}
Assume \textbf{A}.1-\textbf{A}.3. The sequence $(\psi_n, \theta_n)$ converges a.s. to one of the following limits:
\begin{enumerate}[(i)]
    \item $(0,0)$, i.e., both population types get extinct, 
    
    \item $(m-1, 0)$, i.e., only $y$-population survives, 
    \item $(m-1, m-1)$, i.e., only $x$-population survives, or
    \item $(m-1, \tL )$, i.e., both populations co-exist, where
\end{enumerate}
\begin{equation}
    \label{Eqn_tL}
 \tL   :=   (m-1) \bL \mbox{ \normalsize with } \bL := \frac{m_{yx}^*}{m_{xy}^* + m_{yx}^*}.  \hspace{10mm} \mbox{ \eop}
\end{equation} 
\end{theorem}

In Theorem \ref{thrm_high_sep}, we showed that co-existence is not a possibility under high disparity in initial population sizes. While, the above Theorem implies that co-existence is a possibility as well (though, we still do not know the probability of co-existence events and is a part of our future work). Further it is clear that, 
{\it in co-survival sample paths the proportion, $X_n/S_n = \theta_n/\psi_n$, converges to unique    $\bL$ given by \eqref{Eqn_tL}.}

\subsection{\textbf{Individual populations - Growth and Dichotomy}}
 
By Theorem \ref{growth_total_pop}(a), $S_n$ grows at asymptotic rate $\alpha$ and by Theorem \ref{thrm_co_exist}, we have $X_n/S_n$ converges to $0, 1$ or $\bL$, therefore, we get that $x$-type population also grows at rate $\alpha$. Thus, the growth rate of individual populations  with attack  (and  with acquisition also) equals  that without attack. 

Further, on sample paths where $\theta_n \to 0$, $x$-type population becomes minimum population $V_n$ as time progresses. By Theorem \ref{growth_total_pop}(c), it converges to $0$. Thus, {\it $x$-type population either converges to 0 or grows exponentially large, i.e., it exhibits dichotomy}.  Similar is the case with  $y$-type.

{\bf Survival probabilities:}
Now assume that the process starts with equal initial populations, i.e., $x_0 = y_0$ and $\xi_{xy} \stackrel{d}{=} \xi_{yx}$, with $x_0$ not very big. Then by symmetry, the probability of survival of one  of the population types (say that of $x$-type) is more than half that of the total population, $(1- {q^*}^{2 x_0} )/2$ (more  is possible due to  co-existence). However, observe that the probability of survival of either population without attack would have been much larger, as $(1-{q^*}^{x_0}) > (1- {q^*}^{2x_0 } )/2$. We compare the extinction probabilities of two models i.e., BPA and BPNA through simulations in the next subsection.


\subsection{\textbf{Numerical Examples }}
For further insights into our process, we simulate few instances, using Monte-Carlo (MC) simulations. We consider the following parameters: \\
Let the wake-up times be exponentially distributed with parameter, $\lambda = 0.0002$; $F \stackrel{d}{\sim} Poisson(4)$; $\xi_{x}, \xi_y \stackrel{d}{\sim} Bin(F, 0.2667)$; $ \xi_{xy}, \xi_{yx} \stackrel{d}{\sim} Bin(F, 0.053)$. Let the resistance from attack by each population, i.e., $p_{xy}, p_{yx}$, be $0.3$.  
Thus, we get $m = m_x = m_y = 1.067, m_{xy}^* = m_{yx}^* = 0.064$.

Our aim is to compare the extinction probabilities in BPA and BPNA till a pre-defined time,    $T$;  towards this,   we respectively define the extinction probability of  total population, individual populations,  and probability of co-existence of both the population types as:
$q^T_s := P(X_t + Y_t = 0 \mbox{ for some } t \leq T)$, $q^T_x := P(X_t = 0 \mbox{ for some } t \leq T),$ $p^T := P(X_t > 0, Y_t > 0 \mbox{ for all } t \leq T)$.
Similarly define $q^T_y$. We simulate our process till $T = 10^7$ for 3200 instances of MC simulations and  tabulate the estimates of the above probabilities   in Table \ref{table_1}. We set  $y_0 = 2$  and  vary   $x_0$.  

\vspace{-2mm}
\begin{table}[htbp]
\scalebox{0.9}{
\begin{tabular}{|c|c|c|c|c|c|c|c|c|}
\hline
\multicolumn{1}{|c|}{} & \multicolumn{4}{c|}{With attack} & \multicolumn{4}{c|}{Without attack} 
\\ \hline
$x_0$ & $q^T_s$ &$q^T_x$ & $q^T_y$ & $p^T$ & $q^T_s$ &$q^T_x$ & $q^T_y$ & $p^T$ \\ \hline
2    &  0.589 & 0.792 & 0.797  & 0.000  & 0.592 & 0.768 & 0.775  & 0.049        \\ \hline
4   & 0.451 & 0.609 & 0.842  & 0.000   &  0.453 & 0.590 & 0.757  & 0.106        \\ \hline
10   & 0.204 & 0.287 & 0.917 & 0.000 & 0.203 & 0.261 & 0.773  & 0.169        \\ \hline
16 & 0.094 & 0.123 & 0.970  & 0.000 & 0.092 &  0.119 & 0.785 & 0.188 \\ \hline
30   &  0.016 & 0.020 & 0.996  & 0.000  &  0.013 & 0.018 & 0.766 & 0.228   \\ \hline
\end{tabular}}
\caption{\label{table_1}Symmetric  case, with $y_0 = 2$.}
\vspace{-4mm}
\end{table}

From the Table \ref{table_1} we observe that, as illustrated in Theorem \ref{thrm_high_sep},   the  $y$-population gets extinct with increasing probabilities as $x_0$ increases. Further co-existence probability equals zero for all the cases of Table \ref{table_1}.  Observe here that    $m, x_0, y_0$ are all small values.
For the case without attack, i.e.,   for BPNA, as $T \to \infty$,
$$q_s^T, q_x^T, q_y^T  \to   (q^*)^{x_0+y_0}, (q^*)^{x_0},  (q^*)^{y_0} \mbox{ respectively}.$$ 
 We also estimated the same using MC simulations and the results are tabulated in the last four columns of Table \ref{table_1}. The extinction probabilities   of the total population is the same for BPNA as well as BPA, however that of the individual populations are very different from those corresponding to BPA. Further, co-existence is possible (see last column),   by independence, the probability of co-existence  equals $(1-(q^*)^{x_0}) (1-(q^*)^{y_0}).$

\begin{table}[htbp]
    \begin{minipage}{.5\linewidth}
    \centering
    \scalebox{0.9}{
      \begin{tabular}{|c|c|c|c|c|}
\hline
$x_0$ & $q^T_s$ & $q^T_x$ & $q^T_y$ & $p^T$ \\ \hline
200    & 0        & 0.489    & 0.511             & 0                                           \\ \hline
210    & 0         & 0.407     & 0.593              & 0                                           \\ \hline
220   & 0        & 0.319      & 0.681               & 0                                           \\ \hline
250   & 0        & 0.130    & 0.870                & 0                                           \\ \hline
\end{tabular}}
\caption{Symmetric case,\\ with $y_0 = 200$\label{table_2}}
\vspace{-0.4mm}
    \end{minipage}%
    \begin{minipage}{.5\linewidth}
      \centering
      \scalebox{0.9}{
\begin{tabular}{|c|c|c|c|c|}
\hline
$y_0$ & $q^T_s$ &$q^T_x$ & $q^T_y$ & $p^T$ \\ \hline
105    & 0       & 0      & 1            & 0                                           \\ \hline
300    &   0      & 0.023    &  0.977           &       0                                    \\ \hline
4
450    & 0   &    0.548         & 0.452   &  0                                    \\ \hline
500   & 0       & 0.769      & 0.231             & 0    \\ \hline
\end{tabular}}
\caption{ Asymmetric case,\\ with $x_0 = 100$ \label{table_3}}
\vspace{-0.4mm}
    \end{minipage} 
\end{table} 

In Table \ref{table_2}, we consider slightly bigger populations sizes with $y_0 = 200$.  The conclusions are similar to those in the previous Table, but that $q^T_s$ is zero in this case. We also observe that the survival probability of the total population is approximately twice that of the the individual population types (when $x_0 = y_0$). Furthermore, the two population types do not co-exist  even this Table, irrespective of the gaps between initial populations. Observe $m = 1.067$, is small.   More examples are in Tables~\ref{table_5} and \ref{table_6} where we consider co-existence events in both symmetric and asymmetric   cases.

\subsection{\textbf{Asymmetric Case}}
We now  consider  that  one   population dominates in offspring distribution,  i.e., say $\xi_x \stackrel{d}{>} \xi_y$.

\textbf{Growth Patterns:} We additionally require:  
\begin{enumerate}[\textbf{A}.4]
    \item $\inf_{u \geq1} u m_{uv}(v) - v m_{vu}(u) \geq 0, $ where $u$ and $v$ denote the maximum and minimum population respectively.
\end{enumerate}
This assumption implies that the expected  number of attacks by the maximum population is not less than that by the minimum population,  recall $Prob(u\uparrow) = u/(x+y)$. This assumption is satisfied, for example, when $m_{xy}(y)$ is linear in  $y$ or when the number of attacks, $\xi_{xy}, \xi_{yx} \le 1$ a.s. etc. 

\begin{theorem}\label{growth_total_min_asymm_BPDA}Assume \textbf{A}.1-\textbf{A}.2, \textbf{A}.4 and $\xi_x \stackrel{d}{>} \xi_y$. The 
  processes $\{S_n e^{-\alpha_x \tau_n}\}$  and $\{V_n e^{-\alpha_x \tau_n}\}$  are both  non-negative super-martingales, and  converge a.s. and in $L^1$ to  non-degenerate and integrable limits $\widetilde{W}_s$  and  $\widetilde{W}_v$  respectively\footnote{
The proof of Theorem \ref{growth_total_min_asymm_BPDA} is skipped due to lack of space. It follows directly by showing the inequalities of the required conditional expectation, using some coupling arguments. The original system is coupled with BPA where the offsprings of  $y$-type are also distributed like (bigger) $\xi_x$.}. 

\hfill \eop
\end{theorem}

By Theorem \ref{growth_total_min_asymm_BPDA}, we observe that the growth rate of total population is not more than $\alpha_x$.  As in symmetric case,   we also have that the growth rates of the maximum population is  not more than $\alpha_x$ (recall $U_n \le S_n \le 2 U_n$). This gives us that the (asymptotic) growth rates of individual population types is also upper bounded by $\alpha_x$. This is intriguing because {\it such a growth rate (upper bounded by that of BPNA) is observed even when the population types are attacking and acquiring the others and further when the $x$-type has higher reproductive capacity.} One can't comment affirmatively on $y$-type, because these growth rates are upper bounds, i.e., may only satisfy the inequalities like in   \eqref{Eqn_growth_rates}, with only right hand side coefficients as   non-zero values. 

\textbf{Limit Proportions:}
We now proceed towards analysing limit proportions as   in \ref{sec_co_exist} (keeping the notations same). We again have  3-dimensional stochastic approximation based scheme  capturing   $\{\Theta_n\}$  (below, $\xi_{n+1}^y \stackrel{d}{=} \xi_y$ and $\xi_{n+1}^x \stackrel{d}{=} \xi_x$, and rest of the details  as in \eqref{eq_stoch_approx_scheme}):
 
 \vspace{-3mm}
 {\small 
\begin{align}\label{eq_scheme_asymm}
    \psi_{n+1} &= \psi_n + \epsilon_n \bigg( H_n(\xi_{n+1}^x -1)+H_n^c(\xi_{n+1}^y-1) - \psi_n \bigg)K_n,\nonumber \\ 
    \theta_{n+1} &= \theta_n + \epsilon_n \bigg(H_n(\xi_{n+1}^x -1 + \zeta_{xy,n}(Y_n)I_n) \nonumber \\ 
    &\hspace{2cm}- H_n^c\zeta_{yx,n}(X_n)I_n - \theta_n \bigg)J_n.
\end{align}}
The ODE that can approximate \eqref{eq_scheme_asymm} is given by:
{\small 
\begin{equation}\label{ODE_asymm}
    \begin{aligned}
    \dot{\psi} &= \left(\frac{\theta}{\psi}(m_x - m_y) + m_y - 1 - \psi\right)1_{\{\psi > 0 \}},\\
    \dot{\theta} &= \bigg(\frac{\theta}{\psi}\left(m_x-1+m_{xy}((\psi-\theta)\eta(t)) 1_{\{\theta < \psi\}}\right) \\
    &\hspace{1mm}- \left(1-\frac{\theta}{\psi}\right)m_{yx}(\theta \eta(t))1_{\{\theta < \psi\}} - \theta\bigg)1_{\{\psi > 0, \theta > 0 \}}.
    \end{aligned}
\end{equation}}
\DetailK{}{
Further, the ODE corresponding to the ratio $\theta/\psi$ is given as:
{\small \begin{equation}\label{eqn_ODE_theta_psi}
\begin{aligned}
    \dot{\left(\frac{\theta}{\psi}\right)} &=  \frac{1}{\psi}\bigg(
    \frac{\theta}{\psi}
    \left[m_x-m_y + m_{xy}((\psi-\theta)t) 
    + m_{yx}(t\theta)\right] \\
    &-  m_{yx}(t\theta) - \frac{\theta^2}{\psi^2}(m_x-m_y) \bigg) \I ,
    \end{aligned}
\end{equation}}
where $\I:= 1_{\{\psi > 0, \theta > 0, \theta < \psi\}  }.$ Needs to be replaced !!}

One can analyse the above ODEs as   in symmetric case, show that \eqref{ODE_asymm}   approximates \eqref{eq_scheme_asymm} and then derive the asymptotic analysis. Inspired by Theorem \ref{thrm_co_exist}, we conjecture that either both populations get extinct, or  $(\psi_n, \theta_n)$ converges to i) $(m_x-1, m_x-1)$, or ii) $(m_y-1, 0)$, or iii)  $(\psiL, \taL)$, which are zeroes of the RHS of the ODE \eqref{ODE_asymm} (with $\psi > 0, \theta > 0, \theta < \psi$).  Using the ODE for ratio $\theta/\psi$ (as in proof of Theorem \ref{thrm_co_exist}),   $(\psiL, \taL)$ are the zeros of the first RHS term in \eqref{ODE_asymm} and: 

\vspace{-4mm}
{\small \begin{equation*}\label{eqn_theta_psi_asymm}
\begin{aligned}
\frac{(\taL)^2}{(\psiL)^2}(m_x-m_y)  - 
    \frac{\taL}{\psiL}
    \left(m_x-m_y + m_{xy}^* +m_{yx}^*
    \right) -  m_{yx}^*   &= 0.
    \end{aligned}
\end{equation*}}
The zero of the above is given by (the other zero is larger than 1, hence inapplicable):

\vspace{-4mm}
{\small
\begin{align}\label{eq_zero_asymm_theta_psi}
\baL := \frac{\taL}{\psiL} = \frac{1}{2} + \frac{ m_{xy}^* + m_{yx}^* - \sqrt{(m_x-m_y)^2 + (m_{xy}^* + m_{yx}^*)^2}}{2(m_x-m_y)}. 
\end{align}}
And then from \eqref{ODE_asymm},    
$\psiL = m_y - 1 + \baL (m_x - m_y).$
We will {\it validate our conjecture through numerical simulations, while theoretical justification is for future work.}
\begin{remark}
From \eqref{eq_zero_asymm_theta_psi} clearly $\baL  < 1/2$, thus the fraction of $x$-population at equilibrium is smaller. This is interesting given that $x$-type is the dominating one.  When viewed from a different perspective,  for the two populations to co-exist, there should be some sort of balance and that is possible only if {\it the size of the population with bigger potential (reproduction  capacity) is smaller than that  with lower potential.   More surprisingly, this aspect does not depend upon the attack capacity of the individual populations, but only upon the combined capacity $m_{xy}^*+m_{yx}^*$.} 
\end{remark} 


\textbf{Numerical Examples:}
We performed MC simulations for the asymmetric case as well. All the parameters 
are the same as those in Table \ref{table_1}, 
except here, $\xi_{x} \stackrel{d}{\sim} Bin(F, 0.3325),
\xi_{xy} \stackrel{d}{\sim} Bin(F, 0.0667)$. Clearly under asymmetric case, $x$-type population has more potential to grow larger and thus attack $y$-type population. Similar observations are captured in Table \ref{table_4} where the extinction probability of total and $x$-type population has reduced significantly in comparison to symmetric case (see Table \ref{table_1}); while $y$-type population gets extinct with higher probability as $x_0$ increases.
\begin{table}[htbp]
\centering
\scalebox{0.9}{
\begin{tabular}{|c|c|c|c|c|c|c|c|c|}
\hline
\multicolumn{1}{|c|}{} & \multicolumn{4}{c|}{With attack} & \multicolumn{4}{c|}{Without attack} 
\\ \hline
$x_0$ & $q^T_s$ &$q^T_x$ & $q^T_y$ & $p^T$ & $q^T_s$ &$q^T_x$ & $q^T_y$ & $p^T$ 
\\ \hline
2    & 0.218       & 0.396    & 0.822              & 0   & 0.235 & 0.305 & 0.742 & 0.188                 \\ \hline
6    & 0.022      & 0.041       & 0.981         & 0     & 0.021 & 0.028 & 0.729 & 0.265                        \\ \hline
8    & 0.008         & 0.014    & 0.994            & 0   & 0.007 & 0.008 & 0.706 & 0.294                         \\ \hline
10   & 0.002        & 0.005      & 0.998              &0    & 0.001 & 0.002 & 0.700 & 0.298                \\ \hline
\end{tabular}}
\caption{\label{table_4}Asymmetric case, with $y_0 = 2$}
\vspace{-3mm}
\end{table}

\begin{figure}[htbp]
\vspace{-6mm}
    \centering
    \hspace{-5mm}
    \includegraphics[width = 7cm, height = 5.5cm]{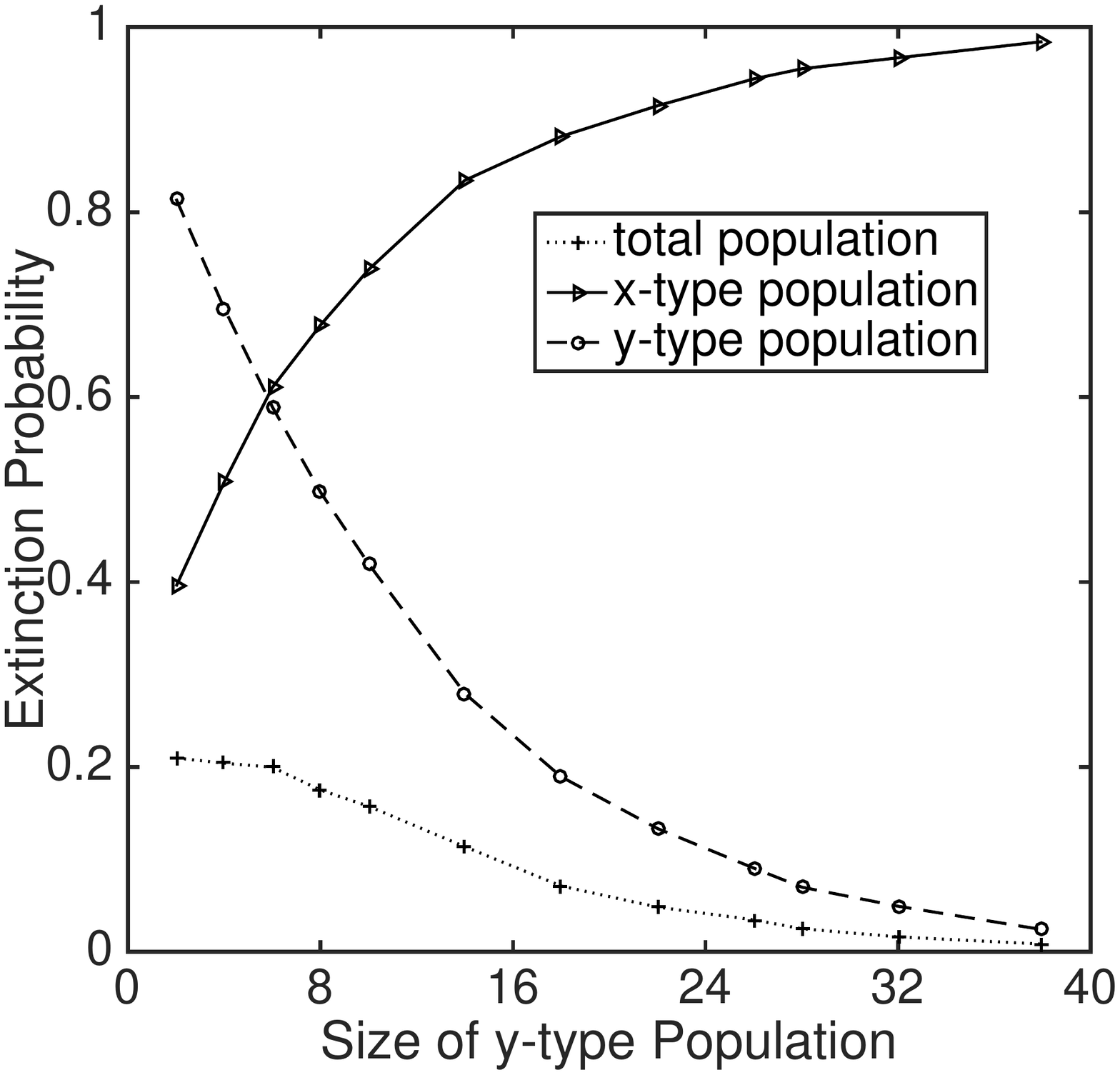}
    \vspace{-9mm}
		\caption{Asymmetric case, with $x_0 = 2$}
		\label{ figure_asymm}
		\vspace{-2mm}
\end{figure}

\begin{table}[htbp]
\centering
\scalebox{0.8}{
\begin{tabular}{|c|c|c|c|c|c|c|c|c|}
\hline
$ 10^{-5}x_0$    & $x_0/y_0$     & $m_x$  & $m_y$  & $ 10^{-5}N$ & $X_N$   & $\bL$ & $X_N/S_N$\\ \hline
1.01 & 1.005      & 2.9998 & 3.0  & 25974       & 2614884763    & 0.501 & 0.503      \\ \hline
10 & 1             & 2      & 2      & 13141         & 659107240    & 0.5   & 0.501      \\ \hline
10     &   0.618    & 3.0 & 2.98        &     10070     & 765240476   & 0.382    & 0.382   \\ \hline
10   &   0.618    & 2.92 & 2.90       &         10534 & 766418907    & 0.382   & 0.381 \\ \hline
\end{tabular}}
\caption{\label{table_5}Co-existence of populations}
\vspace{-1mm}
\end{table}
However, from Figure \ref{ figure_asymm}, we observe that  $x$-type population can also get extinct with large probabilities  when $y_0 > 10$; thus validating Theorem \ref{thrm_high_sep} again.  Further this  is  true even when $m_x + m_{xy}^* > m_y + m_{yx}^*$, i.e., even when the combined reproduction and attack power of $x$-type is bigger than that of $y$-type. 
From Table \ref{table_3} and Figure \ref{ figure_asymm}, we  required $y_0$   approximately 5 times $x_0$ for  significant extinction   of $x$-type population.

\begin{table}[htbp]
\centering
\scalebox{0.9}{
\begin{tabular}{|c|c|c|c|c|c|c|c|}
\hline
& $m_x$ & $m_y$ & $m_c^*$ & $X(0)$ & $Y(0)$ & $\baL$    & $\%_{in}$ \\ \hline
\multirow{4}{*}{Symmetric} & 300   & 300   & 10         & 2500   & 3001   & 0.5      & 11.2,   \\ \cline{2-8}
& 300 & 300 & 10 & 3000 & 3001 & 0.5 & 10.4\\ \cline{2-8}
& 300   & 300   & 10         & 3000   & 3601   & 0.5      & 10.1  \\ \cline{2-8}
& 300   & 300   & 30         & 3000   & 3601   & 0.5      & 5.1  \\ \hline
\multirow{3}{*}{Asymmetric} & 300   & 280   & 10         & 3000   & 8692   & 0.292893 & 7.7  \\ \cline{2-8}
& 300   & 290   & 20         & 3000   & 4611   & 0.438447 & 5.3 \\\cline{2-8}
& 450 & 447 & 20 & 2500 & 3001 & 0.481276 & 7.3 \\ \hline
\end{tabular}}
\caption{\label{table_6} Co-existence with moderate $x_0, y_0$  }
\vspace{-4mm}
\end{table}


Next, we reinforce  co-existence results of Theorem \ref{thrm_co_exist} and support our conjecture for asymmetric case \eqref{eq_zero_asymm_theta_psi} through simulations for the case with {\small $m_{xy}^* = m_{yx}^* = m_c^* = 0.02$} in Table~\ref{table_5}. We see that the limit fractions estimated through MC simulations well match the theoretical ones, \eqref{Eqn_tL} and \eqref{eq_zero_asymm_theta_psi}. Here  $(x_0,y_0)$ are very large. Now we consider slightly bigger values for mean offsprings ($E[F]$  is usually large   in large OSNs) and study the co-existence   probabilities with moderate $(x_0, y_0).$ The results are in Table \ref{table_6}, along with the parameters. Basically we start with $x_0/(y_0+x_0)$ near $\bL$ or $ \baL$ and find the fraction of sample paths in which the ratio converges closer (with $5\%$) to $\bL$ or $ \baL$, after $10^4$ transitions.  We took average over 1000 sample paths. We observe from Table \ref{table_6} that co-existence is possible in considerable number of sample paths. 
\DetailK{}{
\begin{align*}
    \frac{\theta}{\psi} &= \frac{1}{2} + \frac{2 m_c^* + \sqrt{(m_x-m_y)^2 + 4(m_c^*)^2}}{2(m_x-m_y)} \\ 
    &\geq \frac{1}{2} + \frac{2 m_c^* + \sqrt{(m_x-m_y)^2 }}{2(m_x-m_y)} \\ &\geq \frac{1}{2} + \frac{2 m_c^* + (m_x-m_y)}{2(m_x-m_y)} = 1 + \frac{ m_c^*}{(m_x-m_y)} > 1.
\end{align*}}




\vspace{-2mm}
\section{Analysis of Viral markets}\label{sec_viral_markets}

Let $F$ be the number of friends of a typical user which is IID across the  users in the network. A user shares the post (of say $x$-type) to its friends with probability $\eta_x$ depending upon how much he likes the post, which in turn depends upon the attractiveness of the post. The parameter $\eta_x$ is in control of the $x$-CP. In all, we assume that a user with $x$-type ($y$-type) post shares to $Bin(F, \eta_x)$ (resp. $Bin(F, \eta_y)$)
of its friends. 

Among the total number of friends that a $x$-type user decides to share (i.e., $Bin(F, \eta_x)$), there will be three  fractions: \\ (i) a fraction of users who had not received any post yet, these would correspond to offsprings of $x$-type, $\xi_x$; \\  (ii) another fraction  that already received the  $y$-post,  these would correspond to attacked population,  $\xi_{xy}$; and,\\ (iii) a third fraction who already received the  $x$-post.

We model the offsprings corresponding to (i) and (iii) fractions   together as $\xi_x$. We {\it assume that  $\gamma$ (a  small)   fraction\footnote{It would be more realistic to model that this fraction depends upon current live/unread copies of the given post, this would result in population size dependent BPs with attack and this could be a topic of future interest. }   among any given set of users of the network,  will have post of the given type; 
this modeling is reasonable for  huge  networks, 
as the fraction of users with a given   post is not very significant in comparison with the network size.} 

In other words, we assume that, when a user with (say) $x$-post forwards, the expected number of shares to a subset of its friends without any ($x$ or $y$) post equals   $m_x := E[F] \eta_x (1-2\gamma)$, while the expected number of shares to that  subset of its friends with $y$-type post equals  
$ E[F] \eta_x \gamma$.
Among the latter fraction, i.e., the one that receives both the posts, only $p_{xy}$ sub-fraction prefers the new ($x$) post; this preference of the user changes the type of the post to $x$-type and this is {\it exactly equivalent to an attack over $y$-type population by $x$-type.} Thus to model the viral markets using BPA, we assume/have the following modeling details:

\vspace{-4mm}
{\small
\begin{eqnarray}
m_x &=& E[ \xi_x ] = E[ Bin (F, \eta_x(1-2\gamma))] = m_f \eta_x (1-2\gamma), \label{eq_viral_m_x} \nonumber\\
  m_{xy} (y) &=& E[ \zeta_{xy} (y) ] \
  =  \ E[Bin (\min\{\xi_{xy}, y\}, p_{xy}) ]  \\ 
  &=& E[\min\{ Bin (F, \eta_x\gamma), y\}] p_{xy}, \mbox{\normalsize and, } \nonumber \\
  m^*_{xy} &=& \lim_{y \to \infty} m_{xy} (y) = m_f \eta_x \gamma p_{xy }, \xi_{xy} := Bin (F, \eta_x\gamma), \nonumber
\end{eqnarray}}where $m_f := E[F]$.
\textit{The network structure is usually the same towards any post; hence typically the functions $m_{xy}$ and $m_{yx}$ are  (almost) the same (so is $\gamma$). It is $\eta_x$, $\eta_y$ and $x_0, y_0$ that distinguishes the course of the two propagations.}

\noindent{\textbf{BPNA based models}:}
If one models simultaneous  propagation of competing content, without considering attack, 
we will have the two   posts  propagating independent of each other, i.e., then $p_{xy} =   p_{yx} = 0$.  For this case,
$m_x = m_f \eta_x , \mbox{ and } m_y = m_f \eta_y$; basically \textit{the existence of any post (represented by $\gamma$ fraction) does not alter the prospects of any other post}. 
This can be modelled using two independent single type BPs, i.e, BPNAs. 
We compare the conclusions drawn by such independent processes with  those obtained using our BPA model below, to illustrate the drawbacks of the formal models to study   viral competing markets. 

\noindent{\textbf{Some more modeling aspects:}}
Consider two CPs competing for similar product/service.
%
Assume that the posts are stored  in timelines (e.g., as in \cite{dhounchak2017viral}), the inverse stacks on the user's application interface. When a post is shared to the user, this post sits on the top of the timeline and all other posts shift  down by one position. 

Thus, both the type of posts have the potential to attack each other;
 when a user has both the posts it  can prefer the newer post (one on higher levels) to a bigger extent (irrespective of the source of the post). If one of the CPs (say $x$) is more influential than the other, we can model this by $p_{xy} > p_{yx}$. As already mentioned, one designing a more attractive post can be captured by $\eta_x > \eta_y$.  

\subsection*{\textbf{Observations and Comparisons}}
Each CP can control the propagation of their post in two ways: i) by choosing initial seed users (i.e. $x_0, y_0$) and ii) by designing an attractive post (i.e. by appropriately choosing $\eta_x, \eta_y$).  We make the following conclusions:

$\bullet$ \textbf{Significance of number of seed users:}   
Theorem \ref{thrm_high_sep} suggests that irrespective of the quality (or attractiveness) of the post ($\eta_x, \eta_y$), the post which started with higher number of  seeds ($x_0, y_0$) will survive, {\it while the other can get extinct with high probability.} This can also be seen through the simulations in last rows of Tables \ref{table_1} and \ref{table_4}.

However, if instead one models simultaneous  propagation  by {\it two BPNAs,} then we will have two types of posts  propagating independent of each other. Such modelling suggests that {\it the number of seed users of each CP have no affect on the propagation of the other post (see also Table \ref{table_1}, \ref{table_4}); thus providing misleading results about viral competing markets.}

From Theorem \ref{thrm_high_sep}, we can also conclude that if a CP (say $x$-type) invests more in sharing its post to large number of seed users, instead of investing on designing a better post (i.e., $\eta_x$), then irrespective of the the quality of the other post ($y$-type), $x$-CP is always in a better position. If the $y$-CP shares its post to lesser number of seeds, then there is a high possibility that $x$-type post  reaches more users and $x$-type CP can become a monopoly over the network. If not this, then both types of posts  can sustain in the network with $x$-type fraction given by \eqref{Eqn_tL} or \eqref{eq_zero_asymm_theta_psi}. Thus, \textit{designing a better quality post does not highly alleviate the virality chances of the post. Thus, for any CP  the number of seed users is the most critical aspect}, which is not aptly indicated by BPNA based models.

$\bullet$ \textbf {Probability of survival:} 
In viral competing markets, our models  suggest that each post gets extinct with higher probability in comparison to the BPNA based models. Under BPNA, each post (say $x$-type) gets extinct with probability $(q^*)^{x_0}$ (where $q^*$ is defined in Theorem \ref{growth_total_pop}) , which is close to $0$ even when $x_0$ is moderately big   (see  Table \ref{table_1}, \ref{table_4}). In contrast, when one considers attack aspect (initial columns of the Table \ref{table_1}) the extinction probability is higher; this happens because the  BPNA  based models ignore the obvious   impact of existence of competing content. This disparity would be accentuated to a much larger extent in Tables \ref{table_2} and \ref{table_3}; for this case, the extinction probability provided by BPNA based models  ($(q^*)^{x_0}$ or $(q^*)^{y_0}$) is almost zero (comparable to  $q_s^T$ of the Tables), while that derived by attack based models is significantly large.

Additionally, for initial number of seed users as large as the ones   in Tables \ref{table_2} and \ref{table_3}, the two posts can not survive together.  Co-existence of the two posts is possible only  with excessively large number of seed users (as  in Table \ref{table_5}) or in large networks with large $m_f$ (as in Table \ref{table_6}).

$\bullet$ \textbf {Growth rates:}
If we view $\gamma$ (which stands for the fraction of friends of a user having others (competitors) post) as the loss of each CP due to its competitors, then we can see that BPNA does not consider this key characteristic of competing content into consideration. This leads to BPNA suggesting higher $m_x, m_y$ than the ones suggested by equation \eqref{eq_viral_m_x}. %

\section{Conclusions}\label{sec_conclusions}

Online social platforms are usually flooded with variety of content; some of which are competing with each other. The content is stored in an ordered manner as in timelines of Facebook, Twitter etc. Users have tendency to read the content at higher levels and be more lethargic towards the content placed on lower levels, unless the latter belongs to a more influential provider.  A user has a wide variety of choices to choose from, but the competing contents are always at a risk of loosing their chances. 

When a post in a user's timeline gets shifted down by a newer competing post, the new post snatches away the opportunities of (attacks and acquires) the   old post depending upon the popularity and/or the freshness of the two contents.

We propose a \textit{new variant of branching processes} that mimics this `attack' and `acquire'  phenomenon of the competing content. This new variant poses new questions along with the old set of questions (related to growth rates, number of shares and extinction probabilities etc.), that of co-existence of the competing populations. 
In the context of viral competing markets,
this new question translates to co-virality  possibilities, i.e.,  simultaneous 
spread and explosion of the competing content  over the network. 

Our work has two-fold contributions towards the literature related to branching processes, as well as, the social network. The conclusions for branching processes are:\\
$\bullet$ The processes grow at asymptotic rate as without attack.\\ 
$\bullet$ We derived the limit proportions for individual populations using \textit{stochastic approximation techniques}. We showed that either one or both the populations get extinct. If the populations co-exist, they converge to a fixed (unique) ratio. Interestingly, this ratio is smaller for the population with higher reproduction capacity. Further, when the offspring means are different, this ratio depends on combined attack capacity of two populations and not on individual capacities. \\
$\bullet$ We proved that co-existence is not possible when there is a  huge disparity in initial populations. However, it is possible  with large initial populations and/or mean offsprings.

Some of our major conclusions for the social network are: \\
$\bullet$ Number of seed users is the most crucial aspect. Survival chances are  negligible
unless one injects to a large number   of seed users (initial populations). \\
$\bullet$ Probability of simultaneous existence is significantly small, unless one  starts with exorbitantly large number of seed users. This is true even with moderate initial population, when mean number of friends is large  as in huge OSNs.
\\
$\bullet$ One of the caveats of the competition is that the post with smaller number of seed users gets extinct, even after sharing a highly attractive post.\\
Such observations about viral competing markets are not possible (in fact, erroneous conclusions are drawn) when competing contents are modelled as independent branching processes.

\section{Appendix}\label{appendix}

\noindent
\textbf{Proof of Theorem \ref{thrm_high_sep}\footnote{This proof is loosely inspired by the proofs provided in \cite{coffey1991galton}.}:}
We are  considering conditional probability, given that $X_0 = x_0$ and $Y_0 =y_0$ are the respective initial sizes.  
Let $\tau$ be the time epoch before one among the $y_0$ number of particles wakes up; observe that multiple $x$-type particles might have woken up during this time.
Then, $\tau \ge \tau_e$,  where $\tau_e$  is  exponentially distributed random variable with parameter $\lambda y_0$ and equals the  minimum wake-up time among $y_0$ number of $y$-particles.  
Note that some of them could have been attacked/acquired in the meanwhile, and hence $\tau_e$ is only a lower bound. Let $N_{\#}$ be the number of transitions (i.e., wake-ups) of $x$-type before a $y$-type particle wakes up;  then clearly $N_{\#} \geq Bin(x_0, 1-e^{-\lambda \tau})$ a.s.; the lower bound is obtained by considering only those among $x_0$ particles that woke-up. Let $Y'$ denote the size of the $y$-type population at the  next $y$-transition.
Let $P_{0} (\cdot) $ represent the conditional probability $P( \cdot | X_0 = x_0, Y_0 = y_0) $ and similarly let $E_0(\cdot)$ represent the conditional expectation $E( \cdot | X_0 = x_0, Y_0 = y_0) $. 
Now with $\zeta := \zeta_{xy} (1)$ (recall $\zeta_{xy} (1) \le \zeta_{xy}(y)$ a.s. for any $y\ge 1$) we have the following:
%

\vspace{-5mm}
{\small \begin{align*}
    P_0(Y\rightarrow 0)  &\geq P_0(Y' = 0 ) 
     = P_0\Big(y_0<\sum_{i=1}^{N_{\#}} \zeta_{xy,i}\Big)\\
    &\hspace{-14mm}\geq P_0\Big(y_0<\sum_{i=1}^{x_0} \zeta_{i} 1_{\{x \uparrow \}}^\tau \Big) 
    = E_0\Big[P_0\Big(y_{0}<\sum_{i=1}^{x_0} \zeta_{i}1_{\{x \uparrow \}}^\tau \Big |\tau\Big)  \Big],
\end{align*}}where $1_{\{x \uparrow \}}^\tau$  is indicator of the event that the $x$-particle under consideration has woken-up before  $\tau.$ 

Consider a fixed $\overline{ \tau}$ such that (possible as $y_0$ is fixed)
\begin{equation}\label{eq_prob_tau}
    P_0 (\tau \ge {\overline \tau} )>P_0 (\tau_e \ge {\overline \tau} ) = e^{-\lambda y_0 {\overline{\tau}}} > \sqrt{(1- \epsilon)}.
\end{equation}

Let $\Phi $ be the complementary CDF of standard normal random variable  
and  define  the following for the above choice of ${\overline \tau}$:

\vspace{-8mm}
{\small 
$$ \hspace{16mm} Z_{x_0} := \frac{\sum_{i=1}^{x_0} \zeta_{i}1_{\{x \uparrow \}}^{\overline{\tau}} - x_0 m_{xy}(1) (1-e^{-\lambda \overline{\tau}})}{Var(\zeta_{ i})}. $$}
Let $z_{x_0} := \frac{y_0 - x_0 m_{xy}(1)(1-e^{-\lambda \overline{\tau}})}{Var(\zeta_i)}$. Observe that $\{ \zeta_i1_{\{x \uparrow \}}^{\overline{\tau}} \}_i $ are IID random variables and hence by Central Limit Theorem and Portmanteau Theorem for $\varepsilon  := 1-\sqrt{1-\epsilon}$, there exists 
a ${\overline x_0} < \infty$ such that\footnote{Because the Gaussian measure is measure of a continuous random variable, the set  under consideration is Gaussian-continuity set (see \cite{billingsley2013convergence}).} for all $x_0 \ge {\overline x_0}$:
 \vspace{-1.5mm}\begin{eqnarray*}
 P \left (Z_{x_0} > z_{x_0} \right )
 &\ge &
  P_0 \left (Z_{x_0} > z_{\overline{x}_0} \right ) \geq  \Phi \left (  z_{\overline{x}_0} \right ) - \varepsilon/2.
 \end{eqnarray*}
 If required, 
 choose ${\overline x_0}$ further large such that
  \vspace{-1.5mm}
 \begin{equation}\label{eq_prob_cdf}
 P \left (Z_{x_0} > z_{x_0} \right ) \ge  1-\varepsilon \mbox{ \normalsize for all } x_0 \ge {\overline x_0},
 \end{equation}and this is possible because $\Phi (z) \to 1$ when $z \to -\infty$ (hence   $\Phi \left (  z_{\overline{x}_0} \right ) \to 1$ as $\overline{x}_0 \to \infty$).  
 Using the  bounds \eqref{eq_prob_tau}, \eqref{eq_prob_cdf} and conditioning on $\tau$, we have for all $x_0\ge {\overline x_0}$:
 \vspace{-1.5mm}
 {\small 
\begin{align*}
P_0(Y\rightarrow 0) &\geq P_0(Y \to 0; \tau \geq \overline{\tau})\\       &\hspace{-1.7cm}\geq P_0\left( E \left [ z_{x_0}< Z_{X_0} \big |      \tau \right ] ; \tau \geq \overline{\tau} \right) \geq                P_0\left(z_{x_0}< Z_{X_0}\right)P_0(\tau \geq \overline{\tau}) \\
    &\hspace{-1.7cm}= P\left(z_{x_0}< Z_{x_0}\right)P_0(\tau \geq \overline{\tau})\geq  (1-\epsilon). \hspace{3.2cm} \mbox{\eop}
\end{align*} } 

\vspace{-4mm}
\noindent
{\textbf{Proof of Theorem} \ref{growth_total_pop}:}
The first two parts of the theorem are immediate by the discussions above the theorem. \\
\textbf{Part (iii)}
If possible, consider the sample paths  where 
\vspace{-1.5mm}
{\normalsize
\begin{equation}\label{eqn_assumption}
0 < \lim \sup V_n (\omega) < \infty.
\end{equation}}
For each such sample path,   there exists a $M(\omega) >0$, $N(\omega) < \infty$ s.t. $ 0 < V_n \leq M(\omega) \mbox{ for all } n \geq N(\omega).$
As before,  because $S_n/2 \le U_n \le S_n$, we have:

\vspace{-2mm}
{\footnotesize
\begin{align*} 
    P(U_n e^{-\alpha \tau_n} \to 0) &= P(S_n e^{-\alpha \tau_n} \to 0)= P(S_n \to 0) = P(U_n \to 0). 
\end{align*} }
Hence   $W_s(\omega) >0$ (a.s.), because   $U_n \geq V_n > 0 \mbox{ for all } n$. This implies that $ \forall n \geq N(\omega)$, we have that $U_n, V_n$ do not interchange, i.e.,  (WLOG) let $V_n = Y_n  \mbox{ for all such }n $.

Thus, there exists a $\beta > 0$  such that after further increasing $N(\omega)$ (if required), 
$$U_n (\omega)>  \beta e^{\alpha \tau_n(\omega)} \mbox{ \normalsize for all }n \geq N(\omega), \mbox{ \normalsize as } W_s(\omega) >0.$$ 
Observe that\footnote{$\tau_n$ is lower bounded by maximum of $n$-exponential random variables.} $\tau_n  \to \infty$ when $U \nto 0$, and thus $U_n$ increases to $\infty$ for all such sample paths. 
Further as in the proof of  Theorem \ref{thrm_high_sep}, when $y_n \le M(\omega)$, $\exists$ $x_{\epsilon, y_n}$ such that for all $x_n \geq x_{\epsilon, y_n}$:
$$1 \ge P(V \to 0 | U_{n}(\omega)= x_n, V_n(\omega) = y_n) > 1-\epsilon \  \forall \ \epsilon > 0.$$ 
Thus, $V_n \to 0$ in all such sample paths, which implies $\lim \sup V_n = 0$, which contradicts our assumption (\ref{eqn_assumption}).
Thus,   $V_n$ either converges to 0 or $\lim \sup V_n = \infty$. \eop

\vspace{1mm}
\noindent
\textbf{Proof of Theorem \ref{thrm_co_exist}:}
We will prove the result using   \cite[Theorem 2.2, pp. 131]{kushner2003stochastic}, as  ${\bar g}(\cdot)$ is only measurable. Towards this, we first need to prove (a.s.) equicontinuity of sequence $\Theta^n(t) := \Theta_n + \sum_{i=n}^{m(t_n+t)-1}\epsilon_i L_i$, with $m(t) = \eta(t)$. This proof goes through exactly as in the proof of   \cite[Theorem 2.1, pp. 127]{kushner2003stochastic} because of the following reasons:  the random vector  $L_n$ is comprised of $\theta_n, \psi_n$ and IID random variables and by \eqref{Eqn_XnYnSnetc}, it suffices to show that $\sup_n E|\S_n|^2 < \infty$\DetailK{, which is trivially true under {\bf A.}2;}{;
It is easy to observe that ($(x_0, y_0)$ is the initial population) (by (\textbf{A}.2))
\begin{align*}
      E|\S_n|^2 \leq \sup_n \frac{1}{n^2}E|(x_0+y_0)+\sum_{i=1}^n(\xi_i-1)|^2 < \infty.
\end{align*}
Now, we will state Lemmas along with their proof which leads upto the proof of Theorem \ref{thrm_co_exist}.

\begin{lemma}\label{lemma_equi_cont}
Define $\Theta^n(t) = \Theta_n + \sum_{i=n}^{m(t_n+t)-1}\epsilon_i L_i.$
Then, $\{\Theta^n(\omega, \cdot)\}$ is equicontinuous.  \eop
\end{lemma}

\subsection*{\textbf{Proof of Lemma} \ref{lemma_equi_cont}}
The proof of this Lemma can be done in a similar way as done in \cite[Theorem 2.1, PArt 1]{kushner2003stochastic}, while noting that we need to prove equicontinuity for each individual component of $\Theta$. The component-wise analysis of $\Theta_n$ (i.e., individually of $\psi_n, \theta_n, t_n$) is possible because we formulate a separate SA scheme for $t_n$, by considering it as another unknown object. \eop

}
further, we exactly have $E[L_n|{\cal F}_n] = {\bar g} (\Theta_n)$ (here $\beta_n$ in
\cite[Assumption {\bf A}.2.2]{kushner2003stochastic} is 0), as well the projection term $Z_n \equiv  0$. Further, $\{\Theta_n(0)\}_n$ is bounded a.s. by strong law of large numbers as applied to $\{\S_n\}_n$.


In  Lemma \ref{lemma_ASL}, we identify  the  attractors\footnote{ A set $A$ is said to be Asymptotically stable in the sense of Liapunov, if there exist a neighbourhood (called domain of attraction, D($A$)) starting in which the ODE trajectory converges to $A$ as time progresses (e.g., \cite{kushner2003stochastic}). } of \eqref{ODE}, with $\tL, 
\bL$  as in \eqref{Eqn_tL}. 
Proof is now completed  sample-path wise.   

First consider 
the sample-paths in which $\psi_n \to 0$ (i.e., $S_n \to 0$ by Theorem \ref{growth_total_pop}).  Then clearly, $(\psi_n, \theta_n) \to (0, 0)$.
Observe in complementary sample paths, $\psi_n \to m-1$ a.s.

Consider the sample-paths in which  $\{\theta_n\}$  sequence does not exit neighbourhood $\N_\epsilon (\tL) := \{\theta: |\theta-\tL | < \epsilon\}$ infinitely often (i.o.), for every $\epsilon >0$; that is\footnote{From Lemma \ref{lemma_ASL}, $(m-1, \tL, \infty)$    is only an equilibrium point and not an attractor; nonetheless    the actual dynamics can still converge to it.  }, $\theta_n \to \tL$  and so   $(\psi_n, \theta_n) \to (m-1, \tL)$. 
\DetailK{}{
{\color{red}
When $\{\theta_n\}$   exits $\N_{\bar \epsilon} (\tL)$, then $$\theta > \tL + {\bar \epsilon} \mbox{ or } \theta < \tL - {\bar \epsilon} $$
and as $\psi_n$ is converging consider that $\psi_n \in \overline{\N_{\bar \epsilon}}(m-1)$ for all larger $n$. With these two, 
we want  an appropriate $\epsilon >0$ such that, $$\theta/\psi > \bL + \epsilon \mbox{ or } \theta/\psi < \bL - \epsilon.$$ So, when we consider $\theta > \tL + {\bar \epsilon},$ we will get $$\epsilon <  \frac{\tL + \bar{\epsilon}}{m-1+\bar{\epsilon}} - \bL = \frac{(1- \bL) {\bar \epsilon}} {m-1+{\bar \epsilon}} $$ 
Similarly, when $\theta < \tL - {\bar \epsilon},$ we will get $$\epsilon <  \bL  -  \frac{\tL - \bar{\epsilon}}{m-1-\bar{\epsilon}} = \frac{\bar{\epsilon}(1-\bL)}{m-1-\bar{\epsilon}}$$ }}

Now consider the remaining sample paths,   then there exists\footnote{If $\{\theta_n\}$  sequence  exits neighbourhood $\N_{\bar \epsilon} (\tL)$ i.o., and say $\{ \psi_n \}$ entered $\N_{\bar \epsilon} (m-1)$, then choose $\epsilon <   \frac{(1- \bL) {\bar \epsilon}} {m-1+{\bar \epsilon}}$. } at least one $\epsilon >0$ such that  $\{\psi_n, \theta_n, t_n\}$ sequence  visits the compact $S_\epsilon$ of  Lemma \ref{lemma_ASL} i.o. By \cite[Theorem 2.2, pp. 131]{kushner2003stochastic} as applied to these sample paths, the sequence   converges to the attractor $A$ of  Lemma \ref{lemma_ASL}. \eop

\begin{lemma}\label{lemma_ASL} [{\bf Attractors}]
For ODE \eqref{ODE}, 
the set $A := \{(m-1, 0), (m-1, m-1)\} \times \{\infty\}$ is locally asymptotically stable in the sense of Liapunov. For any $\epsilon > 0$, the set\footnote{Define $\overline{\N_\epsilon} (\tL) := \{\theta: |\theta-\tL | \leq \epsilon\}$.}

\vspace{-5mm}
{\small $$S_\epsilon  = \left\{(\psi, \theta): \psi \in \overline{\N_\epsilon}(m-1), \frac{\theta}{\psi} \in [0,1]- \N_\epsilon(\bL) \right\} \times [T_0, \infty],$$}is  compact 
and is in the domain of attraction of $A$, when $T_0$ is such that $$\eta(T_0)  \bL(m-1-\epsilon) \ge \max \left\{ {\bar y}\kappa_{xy}/\kappa_{yx}, \bar{x} \kappa_{yx}/\kappa_{xy}\right\}.$$  Further, the equilibrium point $(m-1, \tL, \infty)$ is not stable. 

\hfill \eop
\end{lemma}
{\bf Proof :} For $t$ component, we consider the extended positive real line including  $\infty$, where the absolute distance metric is naturally extended by $d(t, \infty) := 1/t$; proofs in \cite{kushner2003stochastic} would go through even for this. With this, the required compactness is true.  
The $\psi$-component of the ODE \eqref{ODE} has the following solution:
 
 \vspace{-5mm}
 {\small \[\psi(t) =
\begin{cases}
 e^{-t}(\psi(0) - m + 1)+ m-1,  & \mbox{ \normalsize when } \psi(0) > 0, \\
\psi(0), & \psi(0) \leq 0 .
\end{cases}
\]}
Thus $(m-1)$ is asymptotically stable with $(0, \infty)$ as domain of attraction. For $\theta$ component,  one needs to substitute solution $\psi(t)$ in its ODE (${\bar g}^\theta$ of \eqref{ODE}) to analyze. 
By considering $\psi^* = m-1$, the equilibrium points\footnote{In this context, the point $\bar{\theta}$ is an equilibrium point if ${\bar g}^\theta (\psi^*, {\bar \theta}, \infty) = 0$. }  for the ODE  corresponding to $\theta$ component   are  0, $m-1$ or $\tL.$ 

\DetailK{To test the stability of the above equilibrium points we }{We claim that  equilibrium points  $(m-1, 0, \infty), (m-1, m-1, \infty)$ are stable, while $(m-1, \tL, \infty)$ is unstable. Towards this,
}
 consider the ODE representing the ratio $\theta/\psi$ (derived using~\eqref{ODE} and with $\I:= 1_{\{\psi > 0, \theta > 0, \theta < \psi\}  }$):

\vspace{-4mm}
{\small \begin{align*}
    \dot{\left(\frac{\theta}{\psi}\right)} =  \frac{\I}{\psi}\left\{
    \frac{\theta}{\psi}
    m_{xy}((\psi-\theta)\eta(t)) 
    - \left ( 1 -  \frac{\theta}{\psi} \right )  m_{yx}(\eta(t)\theta)  \right\}  .
\end{align*}}
Consider any $\psi \in \overline{\N_\epsilon}(m-1)$,   $t \ge T_0$  and $\theta / \psi  > \bL$, then by {\bf A}.3 we have:

\vspace{-4mm}
{\small \begin{align*}
\dot{\left(\frac{\theta}{\psi}\right)} &\ge \frac{\I}{\psi}\left\{\frac{\theta}{\psi}\kappa_{xy}\min\{\bar{y}, (\psi-\theta)\eta(t)\} - \left(1-\frac{\theta}{\psi}\right)\kappa_{yx}\bar{x}\right\} \\
&=\frac{\I}{\psi} \min \bigg \{(\bar{y}\kappa_{xy} + \bar{x}\kappa_{yx})\frac{\theta}{\psi} -\bar{x}\kappa_{yx}, \\
& \hspace{22mm}\frac{\theta}{\psi} (\psi - \theta)\eta(t)\kappa_{xy}- \left(1-\frac{\theta}{\psi}\right)\bar{x}\kappa_{yx} \bigg \}   >0.     
\end{align*}}In the above, the first term is positive by \textbf{A}.3 and definition of $\beta^L$ and the second 
one is positive  by  choice of $T_0$, as after dividing the second term by $(\psi-\theta)/\psi$ we get:

\vspace{-4mm}
{\small $$ 
\frac{\theta}{\psi}\psi \eta(t) \kappa_{xy} - \bar{x}\kappa_{yx} >  \bL (m-1-\epsilon) \eta(T_0) \kappa_{xy} - {\bar x} \kappa_{yx} > 0.
$$}Thus the derivative of ratio is positive, and hence the ratio   increases   and     $\theta (t)/\psi(t) \to 1$, when initialized with {\small $\theta(0)/\psi(0) > \bL$.} 
Similarly, when initialized with {\small$\theta(0)/\psi(0) < \bL$}, the derivative 
$\dot{\left(\frac{\theta}{\psi}\right)} < 0$ throughout and  $\theta (t)/\psi(t) \to 0$.

Using the above arguments one can also conclude that from any neighbourhood of $(m-1, \tL, \infty)$, there exist points starting from which the ODE converges either to $(m-1, 1, \infty)$ or $(m-1, 0, \infty)$ and hence the last part.  
\DetailK{}{
Let $h(t):= \left[\frac{m-1+m_{xy} +         m_{yx}}{e^{-t}(\psi(0)-(m-1))+m-1}-1\right]$ and $k(t) : = \theta h(t) - m_{yx}$.
Next, define $V(\theta(t), t) $ as:  $V(\theta(t), t) := -\left(k(t)\right)^2.$ 

Note that $\dot{V} = \pdv{V}{\theta}\pdv{\theta}{t} + \pdv{V}{t}.$ Thus, we get 
\begin{align*}
    \dot{V} &= -2 h(t) k(t) \left[k(t) + \frac{h(t)(\psi(0)-(m-1))}{(m-1+m_{xy}+m_{yx})}\theta e^{-t}\right]
\end{align*}
 Consider $\theta$ in the $\epsilon$-neighborhood of $\theta^*$ such that $\theta \neq \theta^*$.  Choose $t \geq T_0$, where $T_0$ is large enough such that $e^{-t}  \to 0$, thus leading to $\dot{V} < 0.$ Since $V(\theta(t), t) < 0$, therefore, $\theta$ never reaches $\theta^*$. Thus, $(m-1, \theta^*, \infty)$ is not a stable equilibrium point. This proves our claim.}
 \eop
 
\DetailK{}{
\noindent
\textbf{Proof of Theorem \ref{growth_total_min_asymm_BPDA}:}
We need to compute $E[ e^{-\beta (\tau_{n+1} - \tau_n)} |\mathcal{F}_n]$, for some $\beta > 0$ for this Theorem. It can be done as follows:
\begin{align*}
E_n[ e^{-\beta (\tau_{n+1} - \tau_n)}]  
    &=  \int_0^\infty \lambda(X_n + Y_n) e^{-(\beta + \lambda(X_n + Y_n))t} dt \\
    &=\frac{\lambda (X_n + Y_n)}{\lambda (X_n + Y_n) + \beta}.
\end{align*}
\textbf{Part (i)}
By the definition of $\mathcal{F}_n$, $S_n e^{-\alpha_x \tau_n}$ adapts to the increasing sigma algebras $\mathcal{F}_n$.  When $X_n>0, Y_n > 0$, we get:
\begin{align*}
E[S_{n+1} e^{-\alpha_x \tau_{n+1}}|\mathcal{F}_n]  \\
&\hspace{-2.8cm}= e^{-\alpha_x \tau_n}  E\left [ (X_{n+1} + Y_{n+1}) e^{-\alpha_x (\tau_{n+1} - \tau_n)}|\mathcal{F}_n\right ]\\
	&\hspace{-2.8cm}=\frac{\lambda (X_n+Y_n)e^{-\alpha_x \tau_n}}{\lambda (X_n + Y_n) + \alpha_x} \frac{X_n}{X_n+Y_n}  \left ( X_n + m_x - 1 +  Y_n  \right) \\
	&\hspace{-2.5cm}+ \frac{\lambda(X_n+Y_n) e^{-\alpha_x \tau_n}}{\lambda (X_n + Y_n) + \alpha_x}  \frac{Y_n}{X_n+Y_n}   \left ( Y_n + m_y - 1 + X_n \right) \\
	&\hspace{-2.8cm}\leq \frac{(X_n + Y_n)    e^{-\alpha_x \tau_n}  }{\lambda (X_n + Y_n) + \alpha_x}  \left [\lambda (X_n + Y_n )  + \alpha_x \right ]\\
	&\hspace{-2.8cm}\leq (X_{n} + Y_{n})e^{-\alpha_x \tau_n} = S_n e^{-\alpha_x \tau_n}.
\end{align*} 
When $X_n = 0, Y_n = 0$, then the result immediately follows.

Consider another BP with two types of population $x , \hat{y}$ such that $\hat{y}$ has same distribution as that of $x$-type particles of our process and has mean $m_x$. Then using appropriate coupling arguments (i.e., by coupling lifetimes and offspring distributions), for all values of $t$,  $Y(t) \leq \hat{Y}(t)$, hence $Y_n \leq \hat{Y}_n$ at each transition epoch $\tau_n$ corresponding to the original process; and $\hat{S}_n := (X_n + \hat{Y}_n)e^{-\alpha_x \tau_n}$ forms a standard BP. 

Under \textbf{A}.2, $ \hat{S}_n e^{-\alpha_x \tau_n}$ converges in $L^2$. Observe that for all $n$, \vspace{-5mm}
\begin{align*}
    \left(S_n e^{-\alpha_x \tau_n}\right)^2 \leq \left(\hat{S}_n e^{-\alpha_x \tau_n}\right)^2.
\end{align*}
Thus, the proof follows by Martingale convergence theorem \cite{jacod2012probability} for uniformly integrable random variables.

\textbf{Part (ii)}
Clearly, $V_n e^{-\alpha_x \tau_n}$ adapts to the increasing sigma algebras $\mathcal{F}_n$. Consider $E[V_{n+1}e^{-\alpha_x \tau_{n+1}}|\mathcal{F}_n]$ as follows:

$\bullet$ Assume $X_n > Y_n > 0$, then $V_n = Y_n$.  Then, we have:

\vspace{-4mm}
{\small 
\begin{align*}
E[V_{n+1}e^{-\alpha_x \tau_{n+1}}|\mathcal{F}_n ]
    &\leq E\left[Y_{n+1}e^{-\alpha_x \tau_{n+1}}|\mathcal{F}_n \right]\\
    &\hspace{-28mm}= Y_n e^{-\alpha_x \tau_n}E\left[Y_{n+1}e^{-\alpha_x \left(\tau_{n+1}-\tau_n\right)}|\mathcal{F}_n \right]\\
	&\hspace{-28mm}= \frac{(X_n+Y_n) e^{-\alpha_x \tau_n}}{\lambda (X_n + Y_n) + \alpha_x}\frac{Y_n}{X_n+Y_n}\left [\lambda (X_n + Y_n )  + \alpha_y \right ] \\
	&\hspace{-25mm} + \mbox{ \small {$\frac{(X_n+Y_n) e^{-\alpha_x \tau_n}}{\lambda (X_n + Y_n) + \alpha_x}\frac{Y_n}{X_n+Y_n} \left[Y_n m_{yx} (X_n) - X_n m_{xy} (Y_n)\right]$}}\\
	&\hspace{-28mm}\stackrel{a}{<} Y_n e^{-\alpha_x \tau_n} = V_n e^{-\alpha_x \tau_n},
\end{align*}}where inequality `a' holds because $\alpha_y < \alpha_x$ and (\textbf{A}.4) holds.

$\bullet$ When $0 < X_n \leq Y_n$ and $X_n = Y_n = 0$., the proof goes through in almost the same manner.

We know that for all values of $n$, 
\begin{equation}\label{eqn_v_asymm}
    V_n e^{-\alpha_x \tau_n} \leq (X_n + Y_n) e^{-\alpha_x \tau_{n}}.
\end{equation}
The rest of the proof goes through by part (i) and Martingale convergence theorem. \eop
}

\end{document}